\documentclass[conference]{IEEEtran}
\IEEEoverridecommandlockouts
\usepackage{cite}
\usepackage{amsmath,amssymb,amsfonts}
\usepackage{algorithmic}
\usepackage{graphicx}
\usepackage{textcomp}
\usepackage{xcolor}
\usepackage{array, arydshln}
\usepackage{float}
\usepackage{booktabs}
\usepackage{blindtext}
\usepackage{multirow}
\usepackage{multicol}
\usepackage{caption}
\usepackage{tablefootnote}
\usepackage{threeparttable}
\usepackage{comment}
\usepackage{hyperref}
\usepackage{bm}
\usepackage{subcaption}   
\usepackage{url} 
\usepackage{xcolor}
\usepackage{soul}
\sethlcolor{red}

\def\BibTeX{{\rm B\kern-.05em{\sc i\kern-.025em b}\kern-.08em
    T\kern-.1667em\lower.7ex\hbox{E}\kern-.125emX}}
\begin{document}

\title{Enhancing Alzheimer's Detection through Late Fusion of Multi-Modal EEG Features
}

\author{%
  \IEEEauthorblockN{%
    Nguyen Thanh Vinh,\,
    Manoj Vishwanath,\,
    Thinh Nguyen-Quang,\,
    Nguyen Viet Ha,\,\\
    Bui Thanh Tung,\,
    Huy-Dung Han,\,
    Nguyen Quang Linh,\,
    Nguyen Hai Linh,\,
    Hung Cao%
  }
  \thanks{Nguyen Thanh Vinh, Nguyen Viet Ha, and Bui Thanh Tung are with VNU University of Engineering and Technology, Hanoi, Vietnam (e-mail: \{21020710, hanv, tungbt\}@vnu.edu.vn).}%
  \thanks{Manoj Vishwanath and Hung Cao are with University of California, Irvine, California, USA. (e-mail: \{manojv, hungcao\}@uci.edu)}%
  \thanks{Thinh Nguyen-Quang and Huy-Dung Han are with Hanoi University of Science and Technology, Hanoi, Vietnam (e-mail: nqthinh.airesearch@gmail.com; dung.hanhuy@hust.edu.vn).}%
  \thanks{Nguyen Quang Linh and Nguyen Hai Linh are with Central Military Hospital 108, Hanoi, Vietnam. (e-mail: dr.linhnguyenquang@gmail.com; drlinhnguyen108@gmail.com)}%
}

\maketitle

\begin{abstract}

Alzheimer’s disease (AD) is a progressive neurodegenerative disorder characterized by cognitive decline, where early detection is essential for timely intervention and improved patient outcomes. Traditional diagnostic methods are time-consuming and require expert interpretation, thus, automated approaches are highly desirable. This study presents a novel deep learning framework for AD diagnosis using Electroencephalograph (EEG) signals, integrating multiple feature extraction techniques including alpha-wave analysis, Discrete Wavelet Transform (DWT), and Markov Transition Fields (MTF). A late-fusion strategy is employed to combine predictions from separate neural networks trained on these diverse representations, capturing both temporal and frequency-domain patterns in the EEG data. The proposed model attains a classification accuracy of 87.23\%, with a precision of 87.95\%, a recall of 86.91\%, and an F1 score of 87.42\% when evaluated on a publicly available dataset, demonstrating its potential for reliable, scalable, and early AD screening. Rigorous preprocessing and targeted frequency band selection, particularly in the alpha range due to its cognitive relevance, further enhance performance. This work highlights the promise of deep learning in supporting physicians with efficient and accessible tools for early AD diagnosis.

\end{abstract}


\section{Introduction}
\label{sec:introduction}
\IEEEPARstart{A}{lzheimer’s disease (AD)} is one of the most pressing neurodegenerative disorders affecting the elderly population worldwide, particularly individuals over 65 years of age. AD currently affects more than 55 million people worldwide, with projections estimating nearly 152 million cases by 2050 \cite{Breijyeh2020}. Characterized by progressive cognitive decline and memory impairment, AD typically progresses through three stages: in the early stage, patients experience increasing impairment in learning and memory which is accompanied by shrinking vocabulary and reduced word fluency; in the middle stage, further deterioration of memory and language impairs independence and speech difficulties become common; and in the late stage, patients become completely dependent on caregivers, with their vocabulary reduced to simple phrases, single words, or even a complete loss of speech \cite{Forstl1999}.

\begin{figure}[t]
    \centering
    \includegraphics[width=0.85\linewidth]{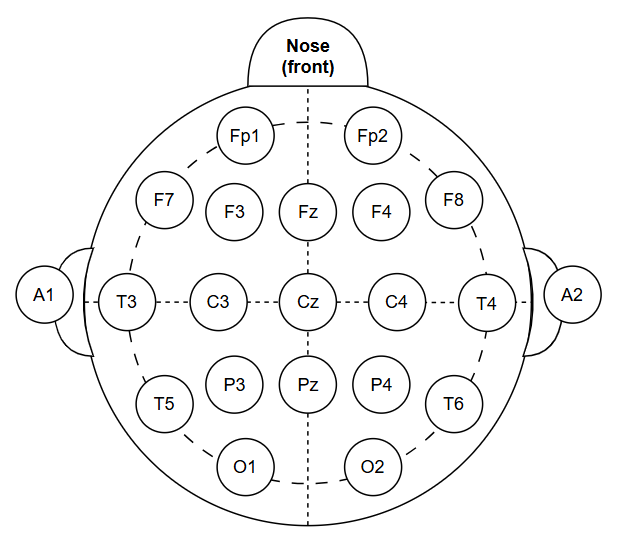}
    \caption{The standard 10-20 electrode placement of EEG recording systems.}
    \label{fig:EEG_10-20_system}
\end{figure}

Electroencephalography (EEG) is a well-established, noninvasive technique for measuring the brain’s electrical activity \cite{amzica2017}. EEG recordings are typically acquired using electrodes placed on the scalp according to the standard 10–20 system or its variations (Figure \ref{fig:EEG_10-20_system}). The EEG signal is composed of multiple frequency bands, each linked to distinct cognitive and physiological functions. Commonly analyzed frequency bands include: Delta (0.1-4Hz), Theta (4-8Hz), Alpha (8-12Hz), Beta (12-30Hz), and Gamma (30-80Hz) \cite{kumar2012}. EEG has played an important role in clinical practice. It has been used to investigate and diagnose a range of neurological disorders, such as epilepsy \cite{vo2022composing, tawhid2022epilepsy, fathima2023baseline}, sleep disorders \cite{cheng2023eegclnet}, stroke, traumatic brain injury \cite{vishwanath2021detection}, and many more, both in humans as well as animal models\cite{benomar2024wireless, vishwanath2020investigation, xia2022microelectrode}. In cognitive neuroscience, EEG plays a key role in studying sensory and auditory processing, emotion recognition \cite{li2022eeg, yin2024multimodal}, memory function, motor control, and general intelligence, due to its excellent temporal resolution and sensitivity to dynamic brain activity. Recent research has also explored the use of EEG for biometric authentication by leveraging inter-subject variability, recognizing the unique neural patterns that distinguish individuals \cite{benomar2022investigation, li2022tensorial}. The study of Alzheimer’s disease using EEG is important because, as the disease progresses, it leads to impaired neural connectivity and changes in EEG such as slowing of alpha rhythms, which can be detected in early stages, before structural damage becomes visible on Magnetic Resonance Imaging (MRI) or Computed Tomography (CT) scans. This makes EEG a promising tool for early detection and diagnosis.

In clinical settings, EEG is commonly used by physicians to assess neurological conditions; however, interpreting these signals manually is time-consuming and requires specialized expertise. To streamline this process, recent research has applied machine learning (ML) \cite{Khatun2019, vecchio2020, Alsharabi2022, hasan2024, gaeta2024, Yu2024} and deep learning (DL) \cite{Kalambe2025, Fouad2023, Wang2025, imani2023} techniques for early detection of Alzheimer’s disease, aiding timely intervention. 
DL approaches offers the potential to automate this process, enhance diagnostic consistency, and uncover subtle patterns in EEG signals that may be difficult for clinicians to detect, thus supporting more efficient and scalable screening of AD in both clinical and remote settings. However, many of these approaches rely on a limited feature set and rarely employ late-fusion strategies, an increasingly promising technique in DL for combining diverse representations to capture complex signal patterns more effectively.

This study proposes a deep learning (DL) model that employs a late-fusion approach, also referred to as decision-level fusion, to classify individuals with Alzheimer's disease from healthy controls using EEG data. In this approach, separate models are trained on different feature sets, and their individual predictions are subsequently merged to produce a final classification. Late fusion \cite{gadzicki2020early, snook2005} is particularly effective for EEG analysis when combining diverse feature representations such as raw time-series signals, Discrete Wavelet Transform (DWT) coefficients, and Markov Transition Field (MTF) images. Each of these captures unique and complementary aspects of EEG data. Time-series EEG signals reflect temporal dynamics, DWT extracts time-frequency patterns, and MTF encodes transition probabilities between signal states as images. By training separate models tailored to each representation, late fusion allows these specialized models to contribute independently learned predictions. Summing the outputs at the decision level enables a more robust and comprehensive classification, leveraging the strengths of each modality while maintaining modularity and scalability. This is especially beneficial in EEG applications where the data is often noisy, high-dimensional, and non-stationary, making a multi-perspective ensemble approach like late fusion ideal for improving generalization and performance.

The remainder of this paper is organized as follows. Section \ref{sec:related_work} reviews related work in the instrumentation and measurement field. Section \ref{sec:method} describes the dataset, data preprocessing steps, feature extraction techniques, and the architecture of the Alzheimer's classifier used in this study. Section \ref{sec:experiments} presents the training procedure and evaluation metrics. Section \ref{sec:results} presents results, and discuss how our work differs in terms of its analytical framework, experimental design, and unique advantages over prior approaches. Section \ref{sec:conclusion} concludes the paper and presents potential future developments.

\begin{table*}[t]
\centering
\caption{Related Works on EEG-Based AD Detection Methods}
\renewcommand{\arraystretch}{1.3} 
\setlength{\tabcolsep}{3pt}       
\begin{tabular}{|
>{\centering\arraybackslash}m{2.5cm}|
>{\centering\arraybackslash}m{0.8cm}|
>{\centering\arraybackslash}m{2.8cm}|
>{\centering\arraybackslash}m{2.2cm}|
>{\centering\arraybackslash}m{5cm}|
>{\centering\arraybackslash}m{0.8cm}|
>{\centering\arraybackslash}m{0.8cm}|
>{\centering\arraybackslash}m{0.8cm}|
>{\centering\arraybackslash}m{0.8cm}|
}
\hline
\textbf{Study} & \textbf{Year} & \textbf{Cohorts} & \textbf{Stimuli} & \textbf{Methodology} & \textbf{AD} & \textbf{FTD} & \textbf{MCI} & \textbf{CN} \\
\hline
Khatun et al. \cite{Khatun2019} & 2019 & 8 MCI, 15 CN & Auditory & Event-related potential and Support Vector Machine with radial basis kernel& - & - & X & X \\
\hline
Vecchio et al. \cite{vecchio2020} & 2020 & 175 AD, 120 CN & Resting State & Brain connectivity, Principle Component Analysis and Support Vector Machine & X & - & - & X \\
\hline
Miltiadous et al. \cite{Miltiadous2021} & 2021 & 10 AD, 10 FTD, 8 CN & Resting State & Decision Tree and Random Forest & X & X & - & X \\
\hline
Alsharabi et al. \cite{Alsharabi2022} & 2022 & 22 AD, 31 MCI, 35 CN & Resting State & DWT and K-Nearest Neighbor & X & - & X & X \\
\hline
Fouad et al. \cite{Fouad2023} & 2023 & 24 AD, 24 CN & Resting State (Eyes Open / Closed) & DWT and Residual Neural Network & X & - & - & X \\
\hline
Hasan et al. \cite{hasan2024} & 2024 & 36 AD, 23 FTD, 29 CN & Resting State & Power Spectrum Density and Linear Discriminant Analysis & X & X & - & X \\
\hline
Yu et al. \cite{Yu2024} & 2024 & 60 AD, 83 CN & Resting State & EEG, genotypes, polygenic risk scores and Support Vector Machine & X & - & - & X \\
\hline
Kalambe et al. \cite{Kalambe2025} & 2025 & 95 AD, 30 FTD, 131 CN & Resting State & Bi-Pyramidal Feature Attention and Separable Networks & X & X & - & X \\
\hline
Stefanou et al. \cite{stefanou2025cnn} & 2025 & 36 AD, 23 FTD, 29 CN & Resting State & Fourier Transform and CNNs & X & X & - & X \\
\hline
\end{tabular}
\label{tab:ad_detection_methods}
\end{table*}

\section{Related Work}
\label{sec:related_work}

In recent years, several studies have used ML and DL methods to automatically classify Alzheimer’s patients. In this section we survey Instrumentation \& Measurement and related literature. ML-based approaches typically involve more careful preprocessing and feature extraction steps, due to the limitations of certain algorithms, such as Naive Bayes and k-nearest neighbors, when handling high-dimensional data such as the EEG signal. 
In \cite{vecchio2020}, the authors used EEG-based brain connectivity and an SVM classifier to distinguish Alzheimer’s patients from healthy individuals. Analyzing network features from 295 subjects, the model achieved high accuracy (AUC 0.97), showing that EEG connectivity can serve as a low-cost, non-invasive tool for early AD detection.
Similarly, \cite{Alsharabi2022} developed an automated EEG-based system for diagnosing AD, where EEG signals were preprocessed and decomposed with DWT. Though it boasts high accuracy, this was obtained using 10-fold cross-validation, unlike our work, which uses a more stringent individual validation approach. In \cite{hasan2024}, the authors focused on spectral features of the EEG signal, estimating the Power Spectral Density (PSD) of EEG recordings and applying Linear Discriminant Analysis (LDA) to discriminate between AD and healthy subjects. In \cite{gaeta2024}, the authors used a combination of clinical variables, conventional polysomnography (PSG) parameters, and quantitative PSG signal features to train ML models to estimate core cerebrospinal fluid biomarkers in Alzheimer’s diagnosis. The results showed that using multimodal input led to a lower Mean Absolute Error compared to using single-feature models.

In contrast, DL-based methods often obviate extensive feature engineering by learning hierarchical representations directly from raw or minimally processed data. In \cite{Kalambe2025}, the authors combined a Bi-Pyramidal architecture and separable networks to extract features and effectively learn representations. In \cite{Fouad2023}, the authors extracted DWT features from EEG signals and fed them into a ResNet architecture, achieving an impressive $97.83\%$ accuracy in classifying Alzheimer's patients. In \cite{imani2023}, the author employed bidirectional long short‑term memory (LSTM) networks and convolutional neural networks (CNN) as separate EEG feature extractors; their outputs were merged via an early‑fusion strategy and passed through a fully connected network, with additional data augmentation performed using an autoencoder. A more recent study introduced LEAD \cite{Wang2025}, the first large foundation model for EEG-based AD detection, built using a large EEG-AD dataset. To tackle limited data and inter-subject variability, the model uses contrastive self-supervised pretraining and channel-aligned unified fine-tuning. LEAD outperforms existing methods and demonstrates the value of large-scale, contrastive learning for robust AD detection. In \cite{mustafa2023}, the authors proposed a multimodal data‐fusion framework for Alzheimer’s diagnosis that combines MRI and CT scan data. Similarly, in \cite{chen2023fusion}, the authors developed a multimodal, feature‑level fusion approach that leverages both time‑ and frequency‑domain EEG inputs to classify Alzheimer’s patients. The results show that integrating multiple modalities yields superior diagnostic performance.

\begin{figure}[t]
    \centering
    \includegraphics[width=1\linewidth]{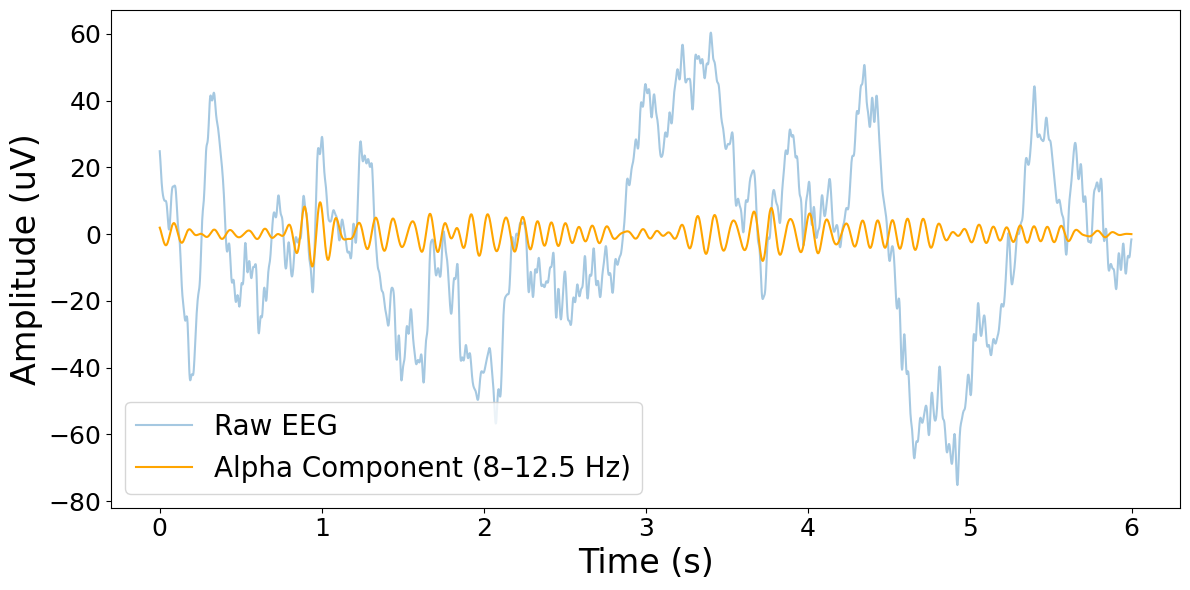}
    \caption{EEG signal before (blue) and after (orange) filtering to extract Alpha-band components.}
    \label{fig:before_after_filtering}
\end{figure}
%
\begin{figure*}[t]
    \centering
    \includegraphics[width=1\textwidth]{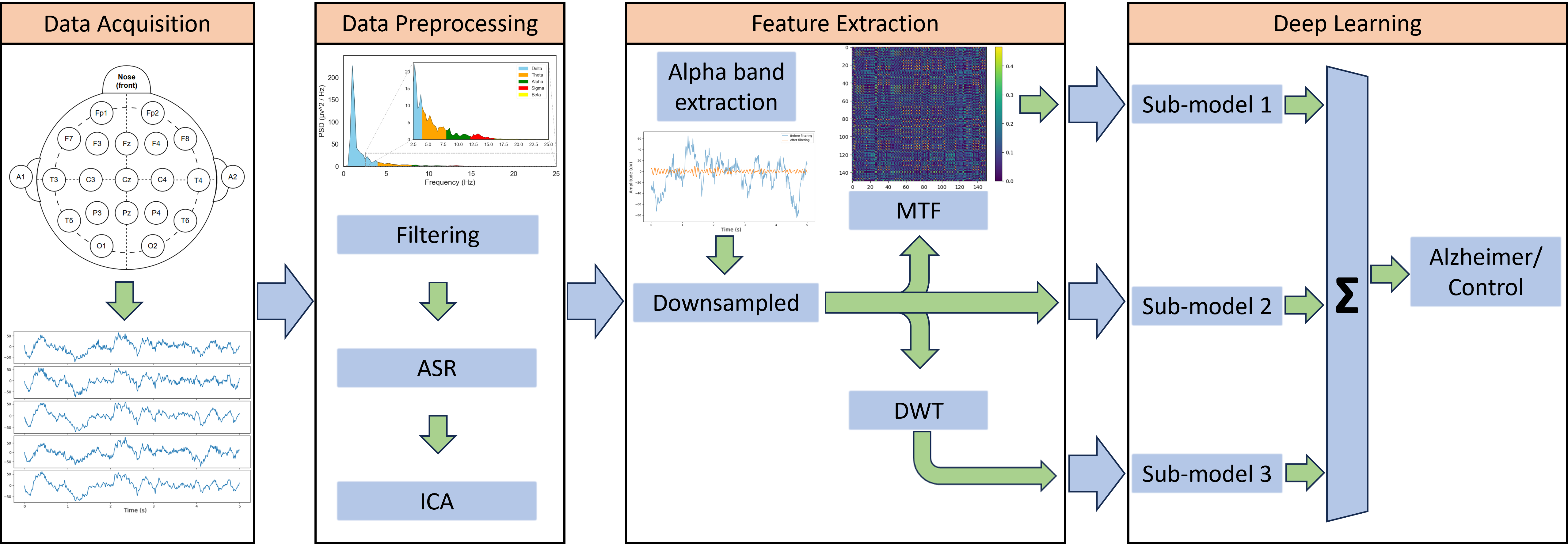}
    \caption{Overall pipeline of the Alzheimer classification based on EEG signal. The EEG data are already preprocessed on the database. We first extract the alpha wave component using a bandpass filter. Next, we perform DWT and MTF transformations. The three resulting feature sets are then fed into three separate models. Finally, the predictions from each model are combined using a late-fusion strategy to produce the final classification result.}
    \label{fig:overall_pipeline}
\end{figure*}

\section{Methods}
\label{sec:method}

\subsection{Dataset}

This study utilizes a publicly available EEG dataset \cite{Miltiadous2024} comprising resting-state, eyes-closed recordings from 88 subjects, categorized into three diagnostic groups: 36 patients with Alzheimer's disease (AD), 23 with Frontotemporal Dementia (FTD), and 29 cognitively normal (CN/HC) controls. Cognitive and neuropsychological state was evaluated by the international Mini-Mental State Examination (MMSE). EEG recordings were acquired at the 2nd Department of Neurology, AHEPA General Hospital of Thessaloniki, using a Nihon Kohden EEG 2100 clinical device based on the 10–20 international electrode placement system, comprising 19 scalp electrodes (Fp1, Fp2, F7, F3, Fz, F4, F8, T3, C3, Cz, C4, T4, T5, P3, Pz, P4, T6, O1, O2) and 2 reference electrodes (A1, A2) placed on the mastoids (Figure \ref{fig:EEG_10-20_system}). Data were sampled at 500Hz. EEG recordings were obtained in a sitting position with eyes closed. The average duration of recordings was 13.5 minutes (range: 5.1–21.3) for AD, 12 minutes (range: 7.9-16.9) for FTD, and 13.8 minutes (range: 12.5–16.5) for HC, totaling 485.5 minutes and 402 minutes for each group, respectively. Table~\ref{tab:demographics} summarizes the demographic and clinical characteristics of the participants.

\begin{table}[t]
\caption{ Demographic and Clinical Characteristics of Participants}
\label{tab:demographics}
\centering
\renewcommand{\arraystretch}{1.05}
\setlength{\tabcolsep}{3.5pt}
\setlength{\extrarowheight}{0.1pt} 
\begin{tabular}{rccc}
\hline\noalign{\vskip 2pt}   
\textbf{Variable} & \textbf{AD (n=36)} & \textbf{FTD (n=23)} & \textbf{CN (n=29)} \\
\hline\noalign{\vskip 2pt}   
Gender (M/F) & 13/23 & 14/9  & 11/18 \\
Age (yrs), mean (SD) & 66.4 (7.9)  & 63.6 (8.2)  & 67.9 (5.4) \\
MMSE, mean (SD)      & 17.75 (4.5) & 22.17 (8.22) & 30.0 (0.0) \\
\hline\noalign{\vskip 2pt}   
\multicolumn{4}{l}{\textbf{Total participants:} 88 (36 AD, 23 FTD, 29 CN)} \\
\hline
\end{tabular}
\end{table}

\subsection{Data Preprocessing}
\label{sec:data_preprocessing}
The preprocessing pipeline included Butterworth band-pass filtering (0.5–45Hz) and re-referencing to the A1–A2 montage, followed by Artifact Subspace Reconstruction (ASR) using EEGLAB to remove high-variance segments based on a conservative threshold (SD = 17). Independent Component Analysis (ICA) was then performed using the RunICA algorithm \cite{delorme2007enhanced}, and components identified as eye or jaw artifacts were automatically removed using ICLabel. The first and last 500 samples were trimmed to eliminate edge noise. The preprocessed data were then divided into 6-second segments and standardized to zero mean and unit standard deviation. Let each preprocessed EEG trial be denoted as $\mathbf{E}_i \in \mathbb{R}^{C \times T}, \ i = 1, \dots, N$, where $N$ denotes number of trial, $C$ is the number of electrodes, $T$ is the number of original time samples.

\subsection{Feature Extraction}
\subsubsection{Time-domain Alpha signal}

The alpha-band component was extracted using a third-order Butterworth bandpass filter (8–12.5Hz) and subsequently downsampled to 25Hz, resulting in trials $\mathbf{A}_i \in \mathbb{R}^{C \times T'}, \ i = 1, \dots, N$, where $T'$ denotes the number of time points after downsampling. A snippet of a trial from electrode Fp1 is shown in Figure \ref{fig:before_after_filtering}. These trials $\mathbf{A}_i$ were then used as input for further feature extraction using MTF techniques. We focused on the alpha-wave band for analysis, as it has been shown to correlate positively with cognitive performance and memory retrieval speed~\cite{bhattacharya2011, rathee2020, lejko2020}.

\subsubsection{Discrete Wavelet Transform}




\begin{figure}[t]
    \centering
    \includegraphics[width=0.75\linewidth]{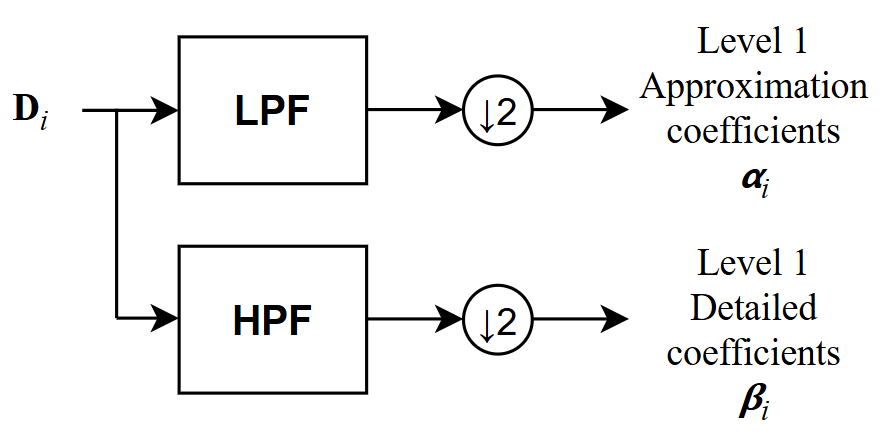}
    \caption{Block diagram of a level‐1 DWT; the approximation coefficients and detailed coefficients are obtained by passing the signal through a low-pass filter and a high-pass filter, respectively, and then downsampled by a factor of 2.}
    \label{fig:level1_DWT}
\end{figure}
The EEG signals $\mathbf{E}_i$ were bandpass-filtered between 8–12.5 Hz to extract the alpha components. Before applying the transformation, we downsampled each trial to 50 Hz, resulting in $\mathbf{D}_i \in \mathbb{R}^{C \times 2T'}$, for $i = 1, \dots, N$.  Wavelet coefficients provide a time–frequency representation of EEG signals, capturing both transient high-frequency events and long-duration low-frequency trends. The DWT of downsampled alpha-band EEG trials $\mathbf{D}_i$ is computed using a two-channel filter bank consisting of a low-pass and a high-pass filter, followed by downsampling. The approximation coefficients, which represent the low-frequency content, are obtained by applying the low-pass filter to $\mathbf{D}_i$ and then downsampling by a factor of two, resulting in $\boldsymbol{\alpha}_i \in \mathbb{R}^{C \times T'}$. Simultaneously, the detail coefficients, which capture the high-frequency components, are derived by filtering $\mathbf{D}_i$ through the high-pass filter followed by downsampling, yielding $\boldsymbol{\beta}_i \in \mathbb{R}^{C \times T'}$. The specific characteristics of the decomposition depend on the chosen wavelet family, which defines the filter coefficients used.
In this study, we use the 16‑tap Daubechies wavelet \cite{Daubechies1992}, whose filter length offers a good compromise between fast convolution/downsampling and adequate frequency resolution. Each EEG trial $\mathbf{D}_i$ was decomposed using a single-level DWT, and the resulting level-1 approximation coefficients $\boldsymbol{\alpha}_i$ were retained as input features for subsequent classification. We opted for a single-level decomposition to balance computational efficiency with information preservation, as higher-level decompositions involve additional downsampling steps that may lead to a loss of information. The DWT was implemented using the PyWavelets library~\cite{PyWavelets}. The 16-tap low-pass and high-pass filter coefficients are:
\begin{equation}
\begin{aligned}
h_{\mathrm{LPF}} &= \{-0.00012,\; 0.00068,\; -0.00039,\;-0.00487,\\
&\quad 0.00875,\; 0.01398,\; -0.04409,\; -0.01737,\\
&\quad 0.12875,\; 0.00047,\; -0.28402,\; -0.01583,\\
&\quad 0.58535,\; 0.67563,\; 0.31287,\; 0.05442\};\\[6pt]
h_{\mathrm{HPF}} &= \{-0.05442,\; 0.31287,\; -0.67563,\;0.58535,\\
&\quad 0.01583,\; -0.28402,\; -0.00047,\; 0.12875,\\
&\quad 0.01737,\; -0.04409,\; -0.01398,\; 0.00875,\\
&\quad 0.00487,\; -0.00039,\; -0.00068,\; -0.00012\}.
\end{aligned}
\end{equation}



\subsubsection{Markov Transition Field}
MTF is a method used to encode time series data into images \cite{Wang2015}.  While spectrograms and wavelets focus on time–frequency content, showing how signal energy is distributed over frequencies and time, MTF captures the temporal dynamics by encoding the transition probabilities between signal states. This makes MTF especially useful for detecting structural patterns and state-dependent behavior in EEG signals. Let each time series downsampled alpha-band EEG trial be denoted as $\mathbf{A}_i \in \mathbb{R}^{C \times T'}, \quad i = 1, \dots, N$, where $C$ is the number of electrodes and $T'$ is the number of time points after downsampling. 
 Assume the values of each channel in $\mathbf{A}_i$ are quantized into $\mathcal{Q}$ bins, and each data point $\mathbf{A}_i(j,k)\ (i = 1, 2, \dots, N;\ j = 1, 2, \dots,C;\ k = 1, 2, \dots, T')$ belongs to only one bin $q_l\ (l = 1, 2, \dots, \mathcal{Q})$. The first-order Markov transition matrix $\mathbf{W}_{\mathcal{Q}\times \mathcal{Q}}$ can then be constructed as follows:
\begin{figure}[t]
    \begin{subfigure}[b]{0.45\textwidth}
        \includegraphics[width=\textwidth]{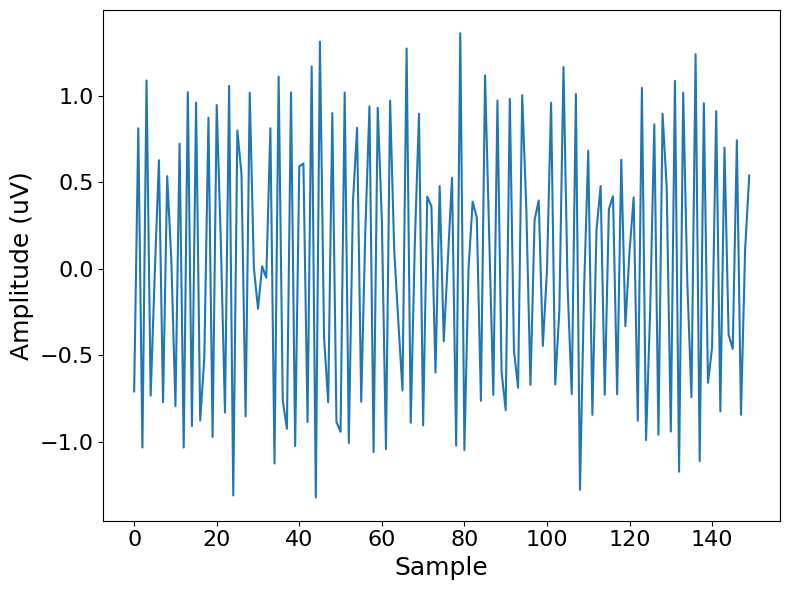}
        \caption{Alpha signal sample}
        \label{fig:timeplot}
    \end{subfigure}
    \hfill
    \begin{subfigure}[b]{0.45\textwidth}
        \includegraphics[width=\textwidth]{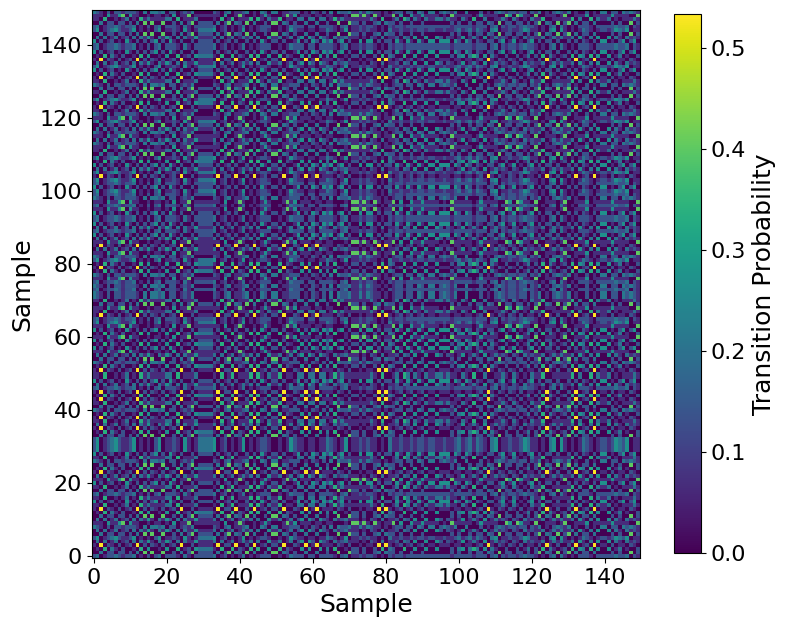}
        \caption{Markov matrix for the sample above}
        \label{fig:markovplot}
    \end{subfigure}
    \caption{MTF visualization: (a) Alpha signal sample; (b) corresponding Markov transition matrix.}
    \label{fig:mtf_eegplots}
\end{figure}

\begin{equation}
   \mathbf{W}_{\mathcal{Q}\times \mathcal{Q}} = \begin{bmatrix}
                    w_{11} & w_{12} & \dots & w_{1\mathcal{Q}}\\
                    w_{21} & w_{22} & \dots & w_{2\mathcal{Q}}\\
                    \vdots & \vdots & \ddots & \vdots\\
                    w_{\mathcal{\mathcal{Q}}1} & w_{\mathcal{\mathcal{Q}}2} & \dots & w_{\mathcal{\mathcal{QQ}}} \\
                 \end{bmatrix}
\end{equation}
where $w_{mn} = p(\mathbf{A}_i(j,k+1) \in q_m\ | \ \mathbf{A}_i(j,k) \in q_n)$ is the probability that the data point $\mathbf{A}_i(j,k+1)$ will belong to bin $q_m$ given that the previous data point $\mathbf{A}_i(j,k)$ belongs to bin $q_n$. Since we are working with a time-series EEG signal, it is evident that each bin follows sequentially from the one preceding it, thus, we have $\sum_{n=1}^{\mathcal{Q}} w_{mn} = 1$. Since each channel consists of $T'$ data points, the output MTF $\mathbf{M}\in \mathbb{R}^{ T' \times T'}$ is defined as follows:
\begin{equation}
   \mathbf{M} = \begin{bmatrix}
        w_{mn} \mid \mathbf{A}_i(j,k) \in q_m,\ \mathbf{A}_i(j,l) \in q_n
    \end{bmatrix}_{k,l=1}^{T'}
\end{equation}
where  $\mathbf{M}(k, l) = w_{mn}\ |\ \mathbf{A}_i(j,k)\in q_m,\ \mathbf{A}_i(j,l)\in q_n,\ (1 \leq i \leq N; 1  \leq j \leq C;1 \leq k, l \leq T' \text{ and }  1 \leq m, n \leq \mathcal{Q}) $ is the probability that $\mathbf{A}_i(j,k)$ of bin $q_m$ at time step $k$ will transition into bin $q_n$ at time step $l$ of $\mathbf{A}_i(j,l)$. By applying this process independently across all $C$ channels, we obtain a final MTF tensor $\bm{\mathcal{M}} \in \mathbb{R}^{C \times T' \times T'}$, where each $T' \times T'$ slice encodes the MTF of one channel. This tensor serves as a multi-channel image-like representation of the EEG signal, which can be used as input for deep learning models. In our experiment, the preprocessed alpha-band EEG time series were converted into images using MTF. The analysis was restricted to four electrode pairs (Fp1–Fp2, O1–O2, P3–P4, and T3–T4), corresponding to the four major brain lobes. We experimented with 5, 10, and 15 quantization bins and found that using 10 bins yielded the best performance. Therefore, the data are quantized into 10 bins ($\mathcal{Q} = 10$) to compute the transition probabilities. The implementation is carried out using the pyts library \cite{pyts}. Figure \ref{fig:markovplot} depicts the results after performing MTF on the downsampled filtered signal shown in Figure \ref{fig:before_after_filtering}.

The overall pipeline is illustrated in Figure \ref{fig:overall_pipeline}. Followed by preprocessing of the raw EEG signal, three distinct input features were calculated: the alpha-band filtered EEG signal, coefficients from the DWT, and 2D MTF. These were fed into three separate sub-models, each inspired by the architecture proposed in \cite{He2016, Nguyen2025}, which is discussed in the next section.

\subsection{Deep learning classifier}
\label{sec:proposed_network}
\begin{figure}[h]
    \centering
    \includegraphics[width=0.9\linewidth]{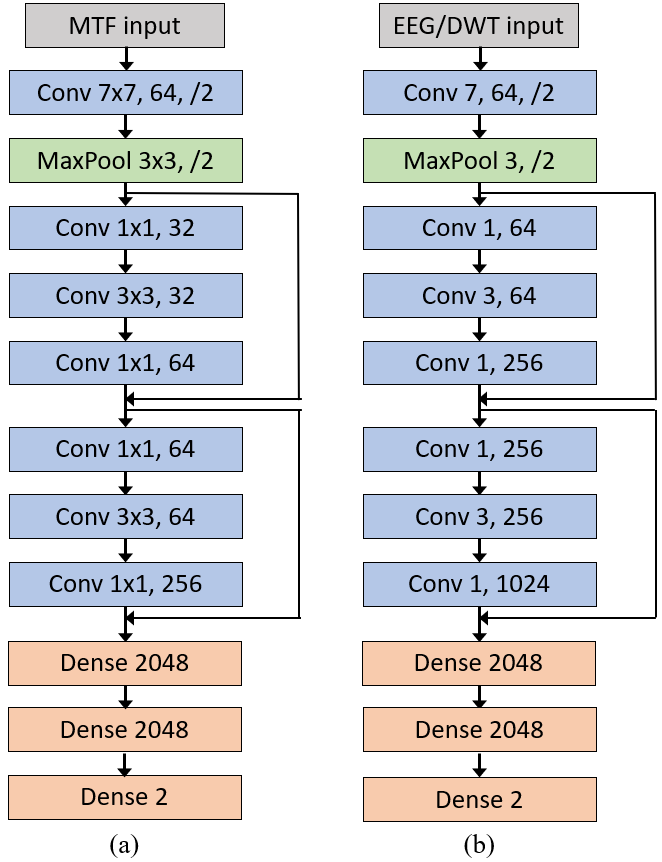}
    \caption{Network architecture of the three sub-models: (a) the sub-model that processes 2D MTF images; (b) the two sub-models that process 1D EEG and DWT inputs. ``Conv'' denotes a convolutional layer and is followed by the kernel size and the number of filters; ``/2'' indicates a stride of 2. If the stride and/or number of filters are not specified, it defaults to 1.}
    \label{fig:DL_network}
\end{figure}

The overall structure of the classifier network is illustrated in Figure \ref{fig:overall_pipeline}. Three CNNs with skip connections are employed to process the alpha-band filtered signal, DWT features, and MTF images independently. The detailed architecture of each sub-model is presented in Figure \ref{fig:DL_network}. Each network consists of convolutional layers with skip connections \cite{He2016}, Batch Normalization \cite{ioffe2015batch}, and Dropout \cite{srivastava2014dropout} layers, and ReLU activation functions \cite{nair2010relu} are applied between layers. The outputs of the three sub-models are integrated using a late-fusion strategy, where their individual predictions are summed to produce the final classification result.

\section{Experiments}
\label{sec:experiments}
\subsection{Training procedure}
\label{sec:experiments/training_procedure}

 The dataset was partitioned into training and test sets as follows. For the AD/HC classification task, 42 patients were randomly selected from a total of 88 subjects for training (21 AD and 21 HC). The test set comprised 16 patients (8 AD and 8 HC) randomly chosen from the remaining pool not included in the training set. The test set was then divided into four equal, patient-based subsets consisting of 2 AD and 2 HC in each scenario. Similarly, for the AD/FTD classification task, 34 patients were randomly selected from a total of 88 subjects for training (17 AD and 17 FTD), and the test set comprised 10 patients (5 AD and 5 FTD) who were not included in training. Each patient-based subset was evaluated independently the mean and standard deviation of each evaluation metric across the four subsets were reported. Table \ref{table:train_class_distribution} lists the number of EEG epochs per class after the segmentation procedure described in Section \ref{sec:method}. Because different recordings yield different numbers of segments, each of the four AD/HC test subsets contains approximately 500 segments per label.

\begin{table}[H]
\centering
\caption{Training set segment distribution}
\begin{tabular}{ccc}
\hline
\textbf{Label} & \textbf{Support} & \textbf{Percentage} \\
\hline
Control & 2921 & 51.71\% \\
Alzheimer & 2728  & 48.29\% \\
\hline
\end{tabular}
\label{table:train_class_distribution}
\end{table}

The proposed network in Section \ref{sec:proposed_network} is trained to minimize cross-entropy loss given by 
\begin{equation}
    \mathcal{L} = -\sum_{i=1}^C y_i\log(\hat{y}_i) + \lambda  \sum_{w_j\in\mathcal{W}} w^2_j
\label{eq:cross_entropy_loss}
\end{equation}
where $w_j$ denotes individual model weight; \( C \) is the number of classes (in our case, \( C = 2 \)); \( y_i \) represents the true one-hot encoded label for class \( i \), where \( y_i = 1 \) for the correct class and 0 otherwise; and \( \hat{y}_i \) denotes the predicted probability of class \( i \), obtained from the softmax function.

The model is trained for 70 epochs on a T4 GPU provided by Google Colab \cite{Google2025} with a total training duration of approximately six minutes using the Adam optimizer \cite{Kingma2015} in the PyTorch framework \cite{Pytorch}. The optimizer's learning rate, $\beta_1$, $\beta_2$, and L2-regularization $\lambda$ are set to 0.001, 0.9, 0.999, and 0.001, respectively.


\subsection{Evaluation metrics}
To evaluate our model, we use $Accuracy$, $Precision$, $Recall$ (or $Sensitivity$), and $F1-score$. Accuracy represents the ratio of correct predictions to the total number of predictions made. Precision and Recall indicate the proportion of true positive predictions among all predicted positives and actual positives, respectively. The F1-score reflects the harmonic mean of precision and recall, providing a balanced measure, especially useful in the presence of class imbalance. 
\begin{equation}
    Accuracy = \frac{TP + TN}{TP + FP + TN + FN}
\end{equation}
\begin{equation}
    Precision = \frac{TP}{TP + FP}
\end{equation}
\begin{equation}
    Recall = \frac{TP}{TP + FN}
\end{equation}
\begin{equation}
    \mathrm{F1\text{-score}} = \frac{2\times Recall \times Precision}{Recall + Precision}
\end{equation}
where \( TP \) = True Positives,  \( TN \) = True Negatives,  \( FP \) = False Positives,  \( FN \) = False Negatives.

\section{Results and Discussion}
\label{sec:results}\

\subsection{Ablation Analysis}

 To assess the contribution of each feature modality, we performed an ablation study that compared different combinations of computed features including unimodal (Table \ref{tab:mtf_channel_pairs}) and multimodal (Table \ref{tab:ablation_results}) fusion models. 

\begin{table*}[!htbp]
\centering
\caption{Ablation results comparing unimodal EEG features.}
\label{tab:mtf_channel_pairs}
\scriptsize
\setlength{\tabcolsep}{6pt}
\renewcommand{\arraystretch}{0.95}
\resizebox{\textwidth}{!}{%
\begin{tabular}{@{}c r l l l l@{}}
\toprule
\textbf{Task} & \textbf{Method} & \textbf{Acc. (+SD)(\%)} & \textbf{Prec. (+SD)(\%)} & \textbf{Rec. (+SD)(\%)} & \textbf{F1 (+SD)(\%)} \\
\midrule
\multirow{10}{*}{\textbf{AD/HC}} 
 & \textbf{(Fp1-Fp2) MTF} & $\mathbf{75.86 \pm 13.36}$ & $\mathbf{76.24 \pm 13.72}$ & $\mathbf{75.69 \pm 13.33}$ & $\mathbf{75.97 \pm 13.52}$ \\
 & \textbf{(O1-O2) MTF}   & $67.70 \pm 12.79$ & $68.93 \pm 13.11$ & $67.78 \pm 12.35$ & $68.34 \pm 12.69$ \\
 & \textbf{(P3-P4) MTF}   & $58.56 \pm 8.49$  & $69.34 \pm 15.43$ & $56.05 \pm 8.06$  & $61.64 \pm 10.85$ \\
 & \textbf{(T3-T4) MTF}   & $73.17 \pm 3.64$  & $74.38 \pm 2.79$  & $72.85 \pm 3.41$  & $73.60 \pm 3.09$ \\
  \cmidrule{2-6}
 & \textbf{Delta}                      & $75.01 \pm 11.66$ & $76.21 \pm 11.35$ & $74.90 \pm 11.72$ & $75.54 \pm 11.52$ \\
 & \textbf{Theta}                      & $76.29 \pm 2.26$  & $78.70 \pm 3.32$  & $75.94 \pm 1.96$  & $77.28 \pm 2.47$  \\
 & \textbf{Alpha}                      & $\mathbf{81.78 \pm 8.38}$  & $\mathbf{82.03 \pm 8.69}$  & $\mathbf{81.66 \pm 8.26}$  & $\mathbf{81.84 \pm 8.47}$  \\
 & \textbf{Beta}                       & $72.31 \pm 15.03$ & $74.31 \pm 14.24$ & $72.92 \pm 14.19$ & $73.60 \pm 14.20$ \\
 \cmidrule{2-6}
 & \textbf{DWT}                        & $\mathbf{81.57 \pm 9.23}$  & $\mathbf{81.86 \pm 9.40}$  & $\mathbf{81.34 \pm 9.30}$  & $\mathbf{81.60 \pm 9.39}$  \\
\midrule
\multirow{10}{*}{\textbf{AD/FTD}} 
 & \textbf{(Fp1-Fp2) MTF} & $\mathbf{68.23 \pm 14.17}$ & $69.44 \pm 12.93$ & $\mathbf{69.33 \pm 13.01}$ & $\mathbf{69.38 \pm 12.97}$ \\
 & \textbf{(O1-O2) MTF}   & $53.83 \pm 3.82$  & $68.24 \pm 4.24$  & $53.51 \pm 1.86$  & $59.95 \pm 2.67$  \\
 & \textbf{(P3-P4) MTF}   & $57.16 \pm 2.18$  & $58.22 \pm 3.34$  & $56.39 \pm 2.53$  & $57.28 \pm 2.90$  \\
 & \textbf{(T3-T4) MTF}   & $61.52 \pm 0.38$  & $\mathbf{70.11 \pm 3.15}$  & $60.69 \pm 2.66$  & $65.06 \pm 2.88$ \\
 \cmidrule{2-6}
 & \textbf{Delta}                      & $68.39 \pm 12.06$ & $69.78 \pm 11.22$ & $68.96 \pm 11.66$ & $69.36 \pm 11.43$ \\
 & \textbf{Theta}                      & $71.96 \pm 3.54$  & $79.12 \pm 3.25$  & $72.43 \pm 3.76$  & $75.62 \pm 3.53$  \\
 & \textbf{Alpha}                      & $\mathbf{77.69 \pm 7.77}$  & $\mathbf{85.08 \pm 3.89}$ & $\mathbf{76.97 \pm 8.09}$ & $\mathbf{80.75 \pm 6.24}$ \\
 & \textbf{Beta}                       & $70.72 \pm 6.35$               & $74.53 \pm 6.70$               & $71.55 \pm 6.20$               & $72.96 \pm 6.16$               \\
 \cmidrule{2-6}
 & \textbf{DWT}                        & $\mathbf{74.03 \pm 5.91}$  & $\mathbf{80.01 \pm 1.86}$  & $\mathbf{74.97 \pm 5.24}$  & $\mathbf{77.36 \pm 3.63}$  \\
\bottomrule
\end{tabular}%
}
\end{table*}

\begin{figure*}[h]
    \centering
    \includegraphics[width=1\linewidth]{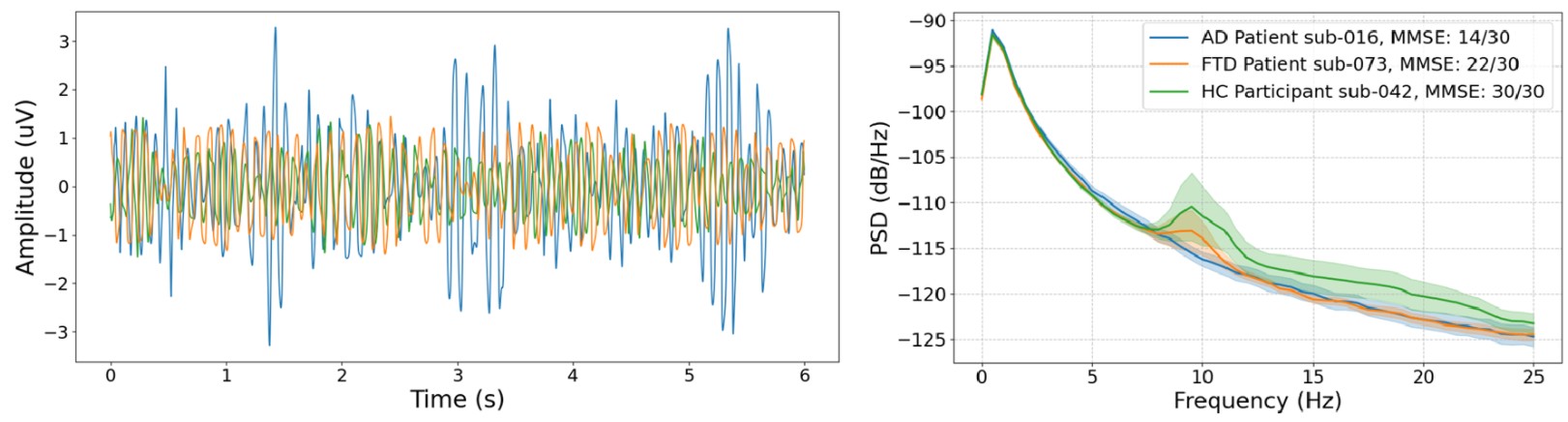}
    \caption{ (Left) Alpha wave signals from channel Fp1 for AD patient (blue), FTD patient (orange) and HC participant (green). It is observed that Alzheimer’s Disease patients exhibit higher alpha wave amplitudes compared to healthy controls and frontotemporal dementia. (Right) Power spectral density (PSD) for AD patient (blue), FTD patient (orange), and HC participant (green). Solid lines show the mean PSD across channels, and shaded areas represent the standard deviation.}
    \label{fig:Fp1_A_and_C}
\end{figure*}

\begin{figure}[htbp]
    \centering
    \includegraphics[width=1\linewidth]{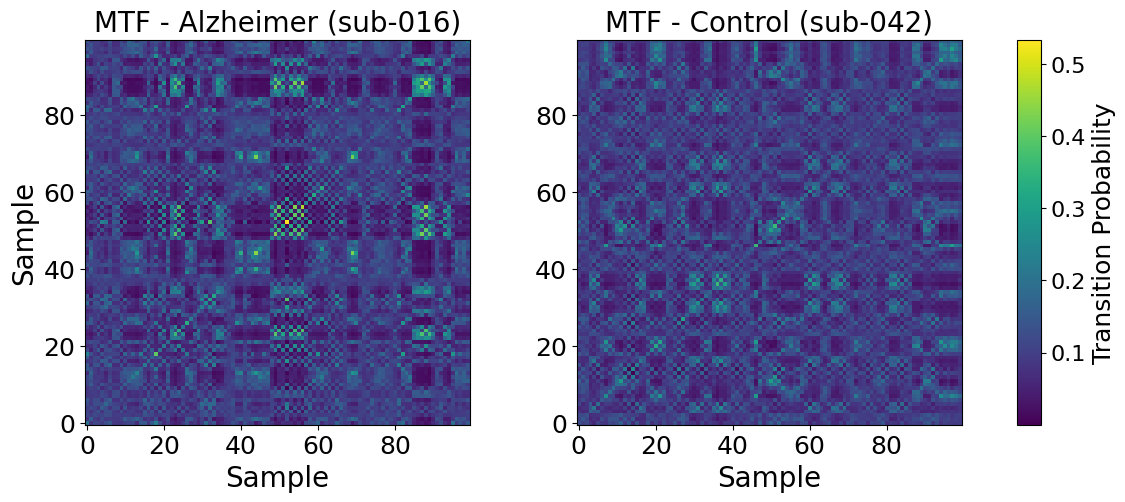}
    \caption{Reduced-size (100$\times$100) MTF representations of the EEG signals shown in  Figure~\ref{fig:Fp1_A_and_C} for an Alzheimer's patient and a healthy control. The MTF images reveal that the Alzheimer's subject exhibits higher transition probabilities in certain states, corresponding to more variability in signal amplitude compared to the healthy control patient.}
    \label{fig:Fp1_A_and_C_mtf}
\end{figure}

In an MTF representation, bright diagonals indicate smooth temporal evolution or high self-transitions, horizontal or vertical streaks suggest persistent states over time, and a checkerboard pattern reflects frequent switching between states, often associated with oscillatory behavior. As shown in Figure~\ref{fig:Fp1_A_and_C_mtf}, the MTF captures differences in signal variability between healthy control and the AD group. In the ablation analysis, we evaluated the MTF-based model under different channel configurations, including Fp1–Fp2, O1–O2, P3–P4, and T3–T4 channels corresponding to the frontal, occipital, parietal, and temporal lobes, respectively (shown in Table~\ref{tab:mtf_channel_pairs}). This analysis aimed to investigate whether MTF representations derived from specific cortical areas can better reflect the pathological network alterations characteristic of AD. As observed in Table~\ref{tab:mtf_channel_pairs}, the frontal pair (Fp1–Fp2) achieved the highest mean accuracy from both classification tasks but was excluded from the final fusion model due to its high variability across test sets. The temporal pair (T3–T4) yielded a more stable performance with a slightly lower mean accuracy and was hence used in the multimodal analysis. The occipital (O1–O2) and parietal (P3–P4) pairs performed much worse on an average. The superior performance of the temporal and frontal pairs can be attributed to their primary functions involving long-term memory \cite{smith2007}, and short-term memory \cite{Pribram1952, Funahashi1993} respectively, both of which are typically impaired in AD and related neurological conditions \cite{kumar2024}. The lower accuracy of the MTF-only model may also be attributable to its use of only two electrodes, a limitation imposed by resource constraints, and this will be addressed in future studies. In contrast, the Alpha and DWT configurations leverage 19 channels, providing a richer spatial representation of brain activity.

\begin{table*}[!htbp]
\centering
\caption{Ablation results comparing multimodal EEG features employing late direct summation fusion.}
\label{tab:ablation_results}
\scriptsize
\setlength{\tabcolsep}{6pt}
\renewcommand{\arraystretch}{0.95}
\resizebox{\textwidth}{!}{%
\begin{tabular}{@{}c r l l l l@{}}
\toprule
\textbf{Task} & \textbf{Method} & \textbf{Acc. (+SD)(\%)} & \textbf{Prec. (+SD)(\%)} & \textbf{Rec. (+SD)(\%)} & \textbf{F1 (+SD)(\%)} \\
\midrule
\multirow{4}{*}{\textbf{AD/HC}} 
 & \textbf{Alpha + DWT}                & $79.50 \pm 14.05$ & $80.45 \pm 13.19$ & $79.38 \pm 14.33$ & $79.90 \pm 13.77$ \\
 & \textbf{Alpha + MTF (T3-T4)}        & $74.57 \pm 13.73$ & $81.93 \pm 8.30$  & $75.47 \pm 12.70$ & $78.43 \pm 10.50$ \\
 & \textbf{DWT + MTF (T3-T4)}          & $83.19 \pm 4.52$  & $84.03 \pm 4.21$  & $83.07 \pm 4.07$  & $83.54 \pm 4.13$  \\
 & \textbf{Alpha + DWT + MTF (T3-T4)}  & $\mathbf{87.23 \pm 6.76}$ & $\mathbf{87.95 \pm 6.75}$ & $\mathbf{86.91 \pm 6.73}$ & $\mathbf{87.42 \pm 6.70}$ \\
\midrule
\multirow{4}{*}{\textbf{AD/FTD}} 
 & \textbf{Alpha + DWT}                & $81.38 \pm 6.76$  & $84.23 \pm 4.55$  & $80.99 \pm 6.92$  & $82.55 \pm 5.74$  \\
 & \textbf{Alpha + MTF (T3-T4)}        & $75.29 \pm 5.40$  & $77.50 \pm 3.34$  & $74.99 \pm 5.60$  & $76.20 \pm 4.48$  \\
 & \textbf{DWT + MTF (T3-T4)}          & $70.22 \pm 5.35$  & $80.78 \pm 2.70$  & $70.07 \pm 4.77$  & $75.02 \pm 3.88$  \\
 & \textbf{Alpha + DWT + MTF (T3-T4)}  & $\mathbf{81.99 \pm 5.18}$ & $\mathbf{84.27 \pm 3.39}$ & $\mathbf{81.67 \pm 5.31}$ & $\mathbf{82.83 \pm 4.33}$ \\
\bottomrule
\end{tabular}%
}
\end{table*}

\begin{table*}[!htbp]
\centering
\caption{Classification performance of AD/HC groups using three EEG features: Alpha, DWT, and MTF (T3-T4) across different fusion strategies.}
\scriptsize 
\label{tab:comparison_of_different_fusion_methods}
\resizebox{\textwidth}{!}{%
\begin{tabular}{@{}rlllll@{}}
\toprule
\textbf{Method} & \textbf{Type} & \textbf{Acc. (+ SD) (\%)} & \textbf{Prec. (+ SD) (\%)} & \textbf{Rec. (+ SD) (\%)} & \textbf{F1 (+ SD) (\%)} \\ 
\midrule
Intermediate fusion & Feature concatenation & $79.53 \pm 12.20$ & $81.58 \pm 10.56$ & $80.04 \pm 11.97$ & $80.79 \pm 11.28$ \\
\midrule
Late fusion & Direct summation & $\mathbf{87.23 \pm 6.76}$ & $\mathbf{87.95 \pm 6.75}$ & $\mathbf{86.91 \pm 6.73}$ & $\mathbf{87.42 \pm 6.70}$ \\
{} & Adaptive weighted & $82.47 \pm 12.85$ & $85.66 \pm 9.38$ & $82.87 \pm 11.92$ & $84.20 \pm 10.69$ \\

\midrule
Multi-classifier & Majority voting & $73.81 \pm 13.12$ & $81.31 \pm 6.63$ & $75.38 \pm 11.75$ & $78.11 \pm 9.40$ \\
{} & Weighted voting & $78.18 \pm 13.93$ & $85.05 \pm 8.12$ & $79.09 \pm 12.77$ & $81.85 \pm 10.57$ \\
\bottomrule
\end{tabular}%
}
\end{table*}

The Alpha time-series model, selected for its positive correlation with cognitive performance, a condition often impaired in Alzheimer’s patients, achieved an accuracy of $81.78 \pm 8.38\%$ for the AD/HC task and $77.69 \pm 7.77\%$ for the AD/FTD task. The improved performance relative to the MTF and DWT models is likely attributable to the distinct differences in alpha-band characteristics between AD patients and HC/FTD patients, as illustrated in Figures \ref{fig:Fp1_A_and_C}.  

The DWT-based model achieved $81.57 \pm 9.23\%$ accuracy for the AD/HC task and $74.03\pm5.91\%$ for the AD/FTD task. Approximation coefficients from the DWT capture the low-frequency, smooth trends of a time series by filtering out fast variations and noise. The magnitude of DWT coefficients reflects the strength of a frequency band in a given time segment, with larger values indicating a stronger presence and smaller values indicating a weaker presence, while the sign of the coefficients indicates the phase of the signal relative to the chosen wavelet.

We observe an improved performance when combining MTF and DWT features, suggesting that the spectral decomposition provided by DWT features complements the temporal transition information captured by MTF. This improvement indicates that while each modality captures distinct characteristics of the EEG signal, their integration enhances the model’s ability to discriminate between AD and control subjects. The most notable performance gain was achieved when Alpha, DWT, and MTF features were used together, underscoring the advantage of leveraging multi-level representations of EEG data for AD classification. Although the Alpha signal alone yielded better performance than the DWT and MTF modalities individually, combining all three feature types in a multi-modal fusion significantly improved the overall classification accuracy, which is attributed to the model’s ability to simultaneously learn and capture the morphological, spectral, and temporal characteristics of the EEG signals.

\subsection{Performance Across Various Fusion Strategies}

\begin{table*}[!htbp]
\centering
\caption{Comparison of the Proposed AD/HC Classification Method with State-of-the-Art Approaches on the Same Dataset}
\scriptsize 
\label{tab:comparison_of_different_methods}
\resizebox{\textwidth}{!}{%
\begin{tabular}{@{}rlllll@{}}
\toprule
\textbf{Work} & \textbf{Method} & \textbf{Acc. (+ SD) (\%)} & \textbf{Prec. (+ SD) (\%)} & \textbf{Rec. (+ SD) (\%)} & \textbf{F1 (+ SD) (\%)} \\ 
\midrule
Miltiadous et al. (2021) \cite{Miltiadous2021} & Decision Tree & $78.50$ & $-$ & $82.40$ & $-$ \\
Miltiadous et al. (2023) \cite{Miltiadous2023} & DICE-Net & $83.28$ & $\mathbf{88.94}$ & $79.81$ & $84.12$ \\
Chen et al. (2023) \cite{chen2023fusion} & CNN + ViT & $\mathbf{87.33}$ & $-$ & $84.56$ & $-$ \\
Hasan et al. (2024) \cite{hasan2024} & PSD + LDA & $76.61 \pm 10.24$ & $80.32 \pm 15.78$ & $69.20 \pm 18.78$ & $72.16 \pm 13.80$ \\
Ma et al. (2024) \cite{ma2024classification} & SVM & $76.90$ & $-$ & $-$ & $-$ \\
Stefanou et al. (2025) \cite{stefanou2025cnn} & FFT + CNN & $79.45 \pm 7.06$ & $76.32$ & $76.06$ & $77.60$ \\
\midrule
\textbf{This paper} & \textbf{Alpha + DWT + MTF} & $87.23 \pm 6.76$ & $87.95 \pm 6.75$ & $\mathbf{86.91 \pm 6.73}$ & $\mathbf{87.42 \pm 6.70}$ \\
\bottomrule
\end{tabular}%
}
\end{table*}

\begin{table*}[!htbp]
\centering
\begin{threeparttable} 
\caption{Comparison of the Proposed Method with State-of-the-Art Approaches}
\scriptsize
\label{tab:comparison_of_different_methods_related_works}
\setlength{\tabcolsep}{10.3pt} 

\begin{tabular}{@{}rllllll@{}}
\toprule
\textbf{Work} & \textbf{Cohorts} & \textbf{Method} & \textbf{Acc. (+ SD) (\%)} & \textbf{Prec. (+ SD) (\%)} & \textbf{Rec. (+ SD) (\%)} & \textbf{F1 (+ SD) (\%)} \\ 
\midrule

Tavares et al. (2019) \cite{tavares2019improvement} & 19 AD / 17 HC & GNN & $95.60$ & $-$ & $-$ & $97.74$ \\
Kamal et al. (2021) \cite{kamal2021alzheimer} & $-$ & CNN + Gene Data & $97.60$ & $-$ & $-$ & $-$ \\
Safi et al. (2021) \cite{safi2021early} & 30 AD / 35 HC & Hjorth Parameters, SVM & $81.00$ & $-$ & $69.80$ & $-$ \\
Klepl et al. (2022) \cite{klepl2022eeg} & 20 AD / 20 HC & SVM, RF, DT, AdaBoost & $92.00 \pm 0.41$ & $-$ & $97.37 \pm 0.94$ & $-$ \\
Lopes et al. (2022) \cite{lopes2023} & 20 AD / 20 HC & CNN, SVM & $87.30$ & $-$ & $-$ & $84.60$ \\
Fouad et al. (2023) \cite{Fouad2023} & 24 AD / 24 HC & ResNet-50 & $97.83$ & $-$ & $-$ & $-$ \\
Wang et al. (2024) \cite{wang2024automatic} & 15 AD / 15 HC & improved AFS + GA & $93.53$ & $-$ & $98.74$ & $-$ \\
\midrule
\textbf{This paper} & 36 AD / 29 HC  & \textbf{Alpha + DWT + MTF} & $87.23 \pm 6.76$ & $87.95 \pm 6.75$ & $86.91 \pm 6.73$ & $87.42 \pm 6.70$ \\
{}& 36 AD / 22 FTD & \textbf{Alpha + DWT + MTF} & $81.99 \pm 5.18$ & $84.07 \pm 3.39$ & $81.67 \pm 5.31$ & $82.83 \pm 4.33$ \\
\bottomrule
\end{tabular}

\begin{tablenotes}
\item[] "$-$" means no data is provided in the existing literature
\end{tablenotes}

\end{threeparttable}
\end{table*}

In this section we discuss different fusion strategies considered in this study. In the intemediate fusion technique, outputs of each model are concatenated after the first Dense layer in Figure \ref{fig:DL_network}. Overall, the intermediate fusion strategy performed worse than late fusion on this dataset, yielding an average accuracy of $79.53 \pm 12.20\%$. For the adaptive weighted method, the outputs from the three submodels were first multiplied element-wise by a learnable weight vector and then passed through a softmax layer, similar to a standard neural network node. After training, the adaptive weights were $\{ \alpha: 0.3944,\ \text{DWT}: 0.3337,\ \text{MTF}: 0.2719\}$, indicating that alpha-band features contributed most strongly to the fused prediction. This result is consistent with the unimodal performance reported in Table \ref{tab:mtf_channel_pairs}, where the alpha model achieved the highest accuracy. We also observed that the adaptive weighted method performed slightly worse than the direct summation fusion. A likely explanation is that element-wise multiplication by weights (which have norms less than one) attenuates the classifier outputs, thereby reducing gradient magnitudes during backpropagation and causing slower convergence relative to direct summation when both methods are trained for the same number of epochs. For the multi-classifier approach, we implemented both majority voting and weighted voting techniques. In majority voting, each model’s prediction contributed one vote, whereas in the weighted voting we assigned votes proportional to unimodal performance $(\alpha:\text{DWT}:\text{MTF} = 5:3:1)$, reflecting the relative accuracies shown in Table~\ref{tab:mtf_channel_pairs}. Weighted voting outperformed majority voting because increasing the influence of the stronger alpha model improved the overall ensemble decision.

\subsection{Comparative Analysis with Existing Methods}

We evaluated our approach using four subsets, as described in Section \ref{sec:experiments/training_procedure}. For the AD/HC classification task, our method achieved an accuracy of $87.23 \pm 6.76\%$, precision of $87.95 \pm 6.75\%$, recall of $86.91 \pm 6.73\%$, and an F1-score of $87.42 \pm 6.70\%$. For the AD/FTD classification task, our method achieved an accuracy of $81.99 \pm5.18\%$, precision of $84.07 \pm 3.39\%$, recall of $81.67 \pm 5.31\%$, and an F1-score of $82.83 \pm 4.33\%$. Table~\ref{tab:ablation_results} shows that the performance on the AD/FTD classification task was slightly lower than on the AD/HC task. This reduction reflects that FTD patients often exhibit EEG characteristics closer to those of AD patients, thereby reducing the separability of the two classes.

Table \ref{tab:comparison_of_different_methods} compares the performance of various Alzheimer’s classification methods on the same dataset, alongside the evaluation metrics achieved by our model. Overall, our approach demonstrates the effectiveness of deep neural networks with late-fusion strategy for AD classification. Compared to traditional machine learning methods such as Decision Tree \cite{Miltiadous2021}, and DL architecture DICE-Net \cite{Miltiadous2023}, which achieved accuracies of $78.50\%$, and $83.28\%$ respectively, our model achieved a higher average accuracy of $87.23\%$. Compared to the CNN–ViT combination in \cite{chen2023fusion}, our approach shows slightly lower performance but offers substantially greater computational efficiency. This improvement in efficiency is largely due to our use of much shorter data segments (6 seconds with downsampling versus 40 seconds), which significantly reduces processing time.

Despite these improvements, the current model's storage requirements ($\sim$ 2 GB) and inference-time memory usage limit its suitability for deployment on typical edge devices, such as Raspberry Pi or EEG headsets equipped with embedded ARM CPUs. Model pruning and quantization (e.g., 8-bit weights, structured pruning) could reduce the size by up to 75\% without significant accuracy loss, as shown in related DL pruning study \cite{han2015}. Additionally, applying knowledge distillation to train a lightweight student model presents a promising direction for enabling real-time, on-device Alzheimer’s screening.

When compared with other state-of-the-art methods, our model sometimes underperforms; however, these comparisons are only indicative because many competing studies used smaller datasets or different modalities (non-EEG), as summarized in Table~\ref{tab:comparison_of_different_methods_related_works}. Moreover, direct comparison is further confounded by a range of methodological differences that affect reported performance. Competing works often employ different preprocessing pipelines (filtering, artifact rejection, channel selection, and segmentation length), distinct feature-extraction or augmentation strategies, and varying numbers of sensors. They may also differ in model class and training procedure (hand-crafted classifiers versus deep networks, pretrained backbones, ensembling, optimization schedules, regularization, and extent of hyperparameter tuning), as well as in class-balancing strategies and the use of data augmentation. Evaluation protocols vary too: some studies report segment-level metrics while others report patient-level results, and cross-validation schemes range from recording-level splits to patient-level held-out tests. Additional sources of discrepancy include differences in diagnostic criteria and label definitions, patient demographics and disease severity, recording conditions (resting vs. task, eyes open/closed), and whether external validation sets were used.

\subsection{Limitations and Future Work}
While our results are encouraging, several limitations warrant discussion. First, the dataset, although publicly available may not fully reflect the demographic diversity and variability in recording hardware encountered in real-world clinical settings (see Table \ref{tab:demographics}). Furthermore, our study focused solely on EEG signals and applied various feature extraction methods, while neglecting other valuable modalities that play an important role in AD classification. Despite improved performance, the current model's storage requirements ($\sim$ 2 GB) and inference-time memory usage limit its suitability for deployment on typical edge devices, such as Raspberry Pi or EEG headsets equipped with embedded ARM CPUs. Model pruning and quantization (e.g., 8-bit weights, structured pruning) could reduce the size by up to 75\% without significant accuracy loss, as shown in related DL pruning study \cite{han2015}. Additionally, applying knowledge distillation to train a lightweight student model presents a promising direction for enabling real-time, on-device Alzheimer’s screening.  Future work will validate our approach on larger, multi-center cohorts, including populations with relevant comorbidities. In addition, longitudinal prediction, such as the progression from mild cognitive impairment to AD, will also be considered.

\section{Conclusion}
\label{sec:conclusion}
In this paper, we proposed a DL pipeline for effective classification of Alzheimer’s patients using late-fusion technique. The results demonstrate that this DL approach outperforms traditional ML algorithms when evaluated on the same dataset. However, the large model size poses a limitation for deployment on edge or wearable devices. In future work, we aim to reduce the model size while maintaining high accuracy to facilitate practical deployment in real-world, resource-constrained environments.


\end{document}